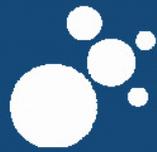

# European Centre for Soft Computing

## Introduction to RIMEP2: A Multi-Expression Programming System for the Design of Reversible Digital Circuits

Fatima Z. Hadjam, Claudio Moraga

Research Report FSC-2014-02

ISSN 2254 - 2736

# Introduction to RIMEP2:
# A Multi-Expression Programming System for the Design of Reversible Digital Circuits

**Abstract**.

Quantum computers are considered as a future alternative to circumvent the heat dissipation problem of VLSI circuits. The synthesis of reversible circuits is a very promising area of study considering the expected further technological advances towards quantum computing. In this report, we propose a linear genetic programming system to design reversible circuits -RIMEP2-. The system has evolved reversible circuits starting from scratch without resorting to a pre-existing library. The results show that among the 26 considered benchmarks, RIMEP2 outperformed the best published solutions for 20 of them and matched the remaining 6. RIMEP2 is presented in this report as a promising method with a considerable potential for reversible circuit design. It will be considered as work reference for future studies based on this method.

## 1. Introduction

Despite their early stage, the reversible circuits are seen as a future alternative to conventional circuit technologies with low-power dissipation applications. In addition, quantum computation also became of interest for reversible circuits since each quantum circuit is inherently reversible. So far, various methods for reversible circuits design have been proposed. Common synthesis approaches and recent comprehensive reviews are presented in [1] and [2].

There are very few publications on the evolutionary design of reversible circuits. Most of them use GAs focusing on finding alternative realization of gates at the quantum level [3] and [4] or on optimizing aspects of already available circuits, like e.g. reordering the outputs [6] and [22]. Works using Genetic Programming are even more rare [5], [23] and [25], although as shown in [7] very good results are obtained.

RIMEP2 was proposed in [7, 8] as Reversible Improved Multi Expression. Optimal competitive solutions were reported for a set of 30 selected benchmarks (n-lines reversible functions, where n=3… 15), where a quantum cost reduction up to 96.13% was reached with an average of 41.08%. The number of ancillary bits was reduced in some cases and new optimal reversible circuits were discovered.

In the following report, we aim to introduce the RIMEP2 system by giving examples on how the chromosome is encoded and how the genetic operations are functioning. The fitness (adaptation function) is fully described. We show a flow chart explaining briefly how RIMEP2 is performing. At the end we will present some of the evolved reversible circuits. A short comparison with some selected published benchmarks is given.

## 2. Introduction to RIMEP2

The synthesis of reversible circuits is more complex than the classical irreversible circuits in a way that additional constraints are added: a reversible circuit has as many inputs as outputs (one-to-one mapping) should be fan-out free (each signal should be used at most once) and cycle free. For more clarity on how a reversible circuit should be specified, an example of the 'AND' logic function before and after reversibility is given by Table 1.

Table 1: The AND logic function. Reversible and irreversible specifications.

| (a) Irreversible truth table | | | (b) Reversible truth table | | | | | |
|---|---|---|---|---|---|---|---|---|
| a | b | f | const | a | b | f | $g_1$ | $g_2$ |
| 0 | 0 | 0 | 0 | 0 | 0 | 0 | 0 | 0 |
| 0 | 1 | 0 | 0 | 0 | 1 | 0 | 0 | 1 |
| 1 | 0 | 0 | 0 | 1 | 0 | 0 | 1 | 0 |
| 1 | 1 | 1 | 0 | 1 | 1 | 1 | 1 | 1 |

According to [17], the minimum number of additional outputs (called, later, garbage) required to convert an irreversible function to a reversible function is $\lceil log_2(M) \rceil$ where M is the maximum number of times that an output pattern is repeated in the corresponding truth table (different input specifications may be used such as truth tables, matrices, decision diagrams [19], Positive Polarity Reed Muller forms [18], cycle-based representations [10], .., etc). Note that in Table 1.a (grey background), the output pattern '0' is repeated 3 times opposing the rule of 'one-to-one mapping' in addition that the numbers of inputs and outputs are different making the AND function specification irreversible, therefore, two additional outputs ($g_1$ and $g_2$) and one constant input (cont) are needed to make the AND function reversible (see Table 1.b).

RIMEP2 is a linear graph-coded genetic programming (GP) approach adapted to the design of reversible circuits. The large searching space and the multi-constraint process to obtain an error free, and fan-out-free reversible circuits of minimum size or minimum quantum cost, make such evolutionary design for reversible circuits a very challenging task. A relevant feature of RIMEP2 is the ability to encode a multi-inputs-outputs reversible circuit within a chromosome compared with a tree-based classical genetic programming which can encode only one output in a chromosome (the root of the encoded tree). RIMEP2 does not, explicitly, encode the inputs and the outputs of an evolved reversible circuit in a chromosome. The outputs are calculated during the evolution.

## 2.1 The chromosome encoding

RIMEP2 chromosome is linear with a pre-fixed length. It consists of a list of expressions (hence the name of the method). Each sub-expression represents a gene and encodes a $k$-bit reversible gate (a reversible circuit is a cascade of reversible gates which implements a given reversible function) with $1 \leq k \leq n$, where $n$ denotes the number of the circuit inputs. A gate may be of several types with different sizes, depending on the library (function set in classical GP) used to evolve the solution of a given problem. Each sub-expression can encode a potential solution (i.e. a reversible circuit). Not necessarily all sub-expressions are involved in the construction of the final solution. To illustrate this encoding, consider the following example: 3-17 from [9]. Its truth table is shown in Table 2.a. RIMEP2 has evolved multiple solutions. Two solutions have been randomly chosen: the first has a quantum cost of 12 with a gate count of 6 and the second has a quantum cost of 11 with a gate count of 5 (explanations about the gate count GC and the quantum cost QC are given in Section 2.2). The corresponding chromosomes are given in Table 2.b. The numbers on the top of Table 2.b do not belong to the chromosome. They indicate only the order of the genes.

Consider the solution 1, the column N° 3 P3(1,2,3). The top of the column indicates the code of the corresponding gate: 'P' where '3' indicates the number of inputs of this gate. The set of gates used to evolve such an example is {P3, T1, T2, T3} encoding respectively Peres, NOT, Feynman (Control Not) and Toffoli (Muti-Control Not) gates (for more information about the gates see for example [9-11]). The remaining scalars represent line numbers. Each line is related to a main input of the reversible circuit being evolved. In this case, a 3-bit reversible circuit is encoded.

In figure 1, the drawings, with a shaded background, represent the effective implementations (only the sub-expressions involved in the implementation of the solution) of the 3-17 problem encoded in the chromosomes (1 and 2 respectively) of the Table 2.b, where the drawings surrounded by dashed line squares represent the complete mappings of the same chromosomes (the whole set of sub-expressions composing the chromosome).

The chromosome is traversed from left to right. The algorithm stops once the full input specification of the searched reversible circuit is met avoiding to pile up additional gates and then, increase the gate count of the evolved solution. For the example cited above, RIMEP2 stopped at the sixth sub-expression in the case of the first solution and at the fifth sub-expression for the second one, thus the corresponding sizes are equal respectively to 6 and 5.

During the evaluation process of the fitness, given a sub-expression, if a line (corresponding to an input of the evolved circuit) is not involved (does not belong to the inputs of the corresponding gate), it continues unchanged (identity). See for example, the first sub-expression of the genotype 1 (line 1 and line 2) in Table 2.b. The gate needs only one input which corresponds to line 3.

RIMEP2 guarantees that any evolved reversible circuit is syntactically correct, cycle and fan-out free simply by avoiding the double use of the same line for a same gate. This is particularly important when a mutation occurs. See section 2.3.

RIMEP2 does not need necessarily a reversible specification of the function to be evolved, thus, to act correctly additional constant lines (called ancillary lines) are added automatically. Some of the outputs (lines) will satisfy the specification of the target circuit. The remaining outputs (needed for reversibility purpose) are not relevant and usually called "garbage".

## 2.2 Fitness evaluation

RIMEP2 evolves guided by a multi-constraints adaptation function (Fitness) $F$: $F(indiv\ i) = (f1, f2, f3)$, where $indiv\ i$ refers to individual $i$ (evolved reversible circuit $i$). $f1, f2$ and $f3$ indicate respectively, the number of errors (the no matching patterns, see Table 3, the '0's and '1's with shaded background), the quantum cost and the gate count of the circuit. The Fitness must be minimized.

**Table 2**

(a) The specification of the 3-17 problem.

| Inputs | | | Outputs | | |
|---|---|---|---|---|---|
| a | b | c | x | y | z |
| 0 | 0 | 0 | 1 | 1 | 1 |
| 0 | 0 | 1 | 0 | 0 | 1 |
| 0 | 1 | 0 | 1 | 0 | 0 |
| 0 | 1 | 1 | 0 | 1 | 1 |
| 1 | 0 | 0 | 0 | 0 | 0 |
| 1 | 0 | 1 | 0 | 1 | 0 |
| 1 | 1 | 0 | 1 | 1 | 0 |
| 1 | 1 | 1 | 1 | 0 | 1 |

(b) The encodings of the evolved solutions.

| 1 | 2 | 3 | 4 | 5 | 6 | 7 |
|---|---|---|---|---|---|---|
| The genotype of the solution 1 | | | | | | |
| T1 | T2 | P3 | T2 | P3 | T2 | T1 |
| 3 | 3 | 1 | 1 | 2 | 1 | 1 |
| - | 2 | 2 | 3 | 3 | 3 | - |
| - | - | 3 | - | 1 | - | - |
| The genotype of the solution 2 | | | | | | |
| T2 | T1 | T2 | P3 | P3 | T1 | T1 |
| 1 | 3 | 2 | 3 | 2 | 1 | 3 |
| 3 | - | 1 | 2 | 1 | - | - |
| - | - | - | 1 | 3 | - | - |

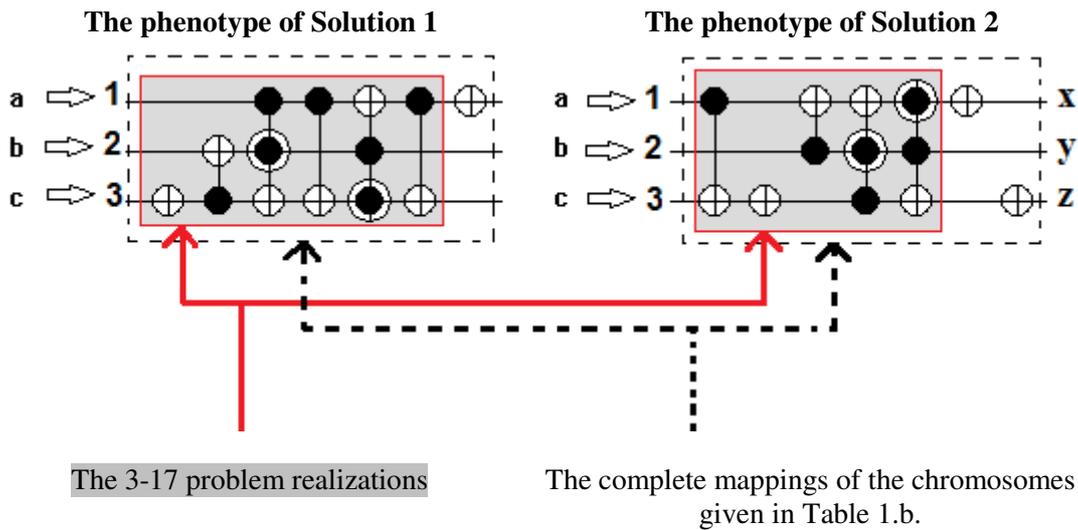

The 3-17 problem realizations

The complete mappings of the chromosomes given in Table 1.b.

**Figure 1:** Two evolved realizations of the 3-17 problem. Gates with a circled dot represent Peres gates.

**Table 3:** The 3-17 problem: The target solution versus the evolved solution.

| Inputs | | | Target Outputs | | | An evolved circuit outputs with $f1 = 8$ | | |
|---|---|---|---|---|---|---|---|---|
| a | b | c | x | y | z | x | y | z |
| 0 | 0 | 0 | 1 | **1** | **1** | 1 | **0** | **0** |
| 0 | 0 | 1 | 0 | 0 | 1 | 0 | 0 | 1 |
| 0 | 1 | 0 | 1 | **0** | 0 | 1 | **1** | 0 |
| 0 | 1 | 1 | 0 | 1 | 1 | 0 | 1 | 1 |
| 1 | 0 | 0 | 0 | 0 | 0 | 0 | 0 | 0 |
| 1 | 0 | 1 | **0** | **1** | **0** | **1** | **0** | **1** |
| 1 | 1 | 0 | **1** | 1 | 0 | **0** | 1 | 0 |
| 1 | 1 | 1 | 1 | **0** | 1 | 1 | **1** | 1 |

Any evolved circuit should first and mainly match the behavioural specification of the target circuit. Two cases may occur:

1. The perfect fit (zero-errors) is met at a given sub-expression. The algorithm stops and considers the ordering number of this sub-expression as the value of *f3*. The cumulated quantum cost until this sub-expression (see the last paragraph of the current section) will be recognized as the quantum cost of the whole evolved reversible circuits and assigned to $f2$.
2. The end of the chromosome is reached and the perfect fit has not been met. The fitness of the chromosome is the fitness of the best sub-expression encoded in this chromosome, i.e., the sub-expression with the minimal $(f1, f2, f3)$. Let $i$ and $j$ be two sub-expressions having respectively $Fi = (fi1, fi2, fi3)$ and $Fj = (fj1, fj2, fj3)$ as fitness. $Fi < Fj$ if:
    a. $fi1 < fj1$ or
    b. $fi1 = fj1$ and $fi2 < fj2$ or
    c. $fi1 = fj1$ and $fi2 = fj2$ but $fi3 < fj3$.

To clarify, consider the chromosome 2 of Table 1.b, RIMEP2 has calculated the fitness of every sub-expression (7 sub-expressions) and are given respectively as follows: (12,1,1), (8,2,2), (10,3,3), (6,7,4), **(0,11,5)**, (8,12,6), (16,13,7). The sub-expression N° 5 with the fitness equal to (0,11,5) is considered to be the best and constitutes the solution (see Figure 1, the drawing with a shaded background of the solution 2).

Each Reversible gate is built utilizing elementary gates where the quantum cost of each is assumed to be 1. The quantum cost of the reversible gate is equal to the number of the elementary gates which constitute it (see [9] and [10]). Examples of elementary gates at the quantum representation are given in Figures 2.b, 3.b and 4.b. **CV**, **CV⁺** and **CNOT** are such cases of elementary gates with a quantum cost of 1. However, some simplifications may be possible. Consider, for example, the references [11] and [12], a kind of post-analysis is realized: looking for groups of NCT gates (building blocks) mainly equivalent to some of the mixed-polarity Peres gates and assigned with a quantum cost of 4. Following Altenberg's Genetic Engineering principle [13], a basic strategy of artificial evolution consists in enlarging the set of basic constructs (gates in this case) with constructs (building blocks) appearing in good solutions, in order to improve the exploration phase of the evolutionary algorithm. Consequently, we did include those mixed-polarity Peres gates in the basic gates library, called NCT which comprises the gates T1, T2 and T3 which have respectively the quantum cost of 1, 1 and 5 according to [9].

For quantum cost calculation purposes:
- The Peres gate has a quantum cost of 4 [14].
- The Peres gate with the 'first' control signal negated has also a cost of 4 [15]. See Figure 2.
- The Peres gate with the 'second' control signal negated is also known as 'inverted' or 'mirror' Peres gate. It has a cost of 4, (see [12] and [15]). See Figure 3.
- An Or-Peres gate has a cost of 4 (see [16]). See Figure 4.

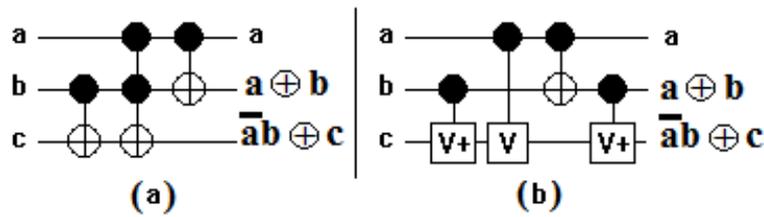

**Figure 2:** The Peres gate with the first control signal negated. (a): The gate level representation. (b): The quantum level representation. V and V+ represent the square root of NOT and its adjoint, respectively.

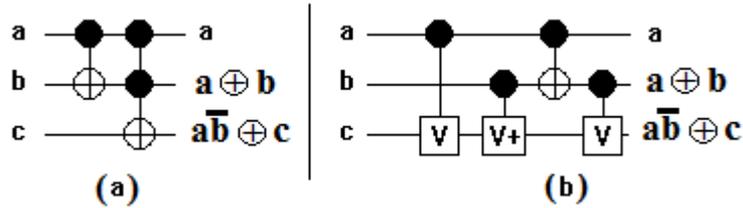

**Figure 3:** The Peres gate with the second control signal negated. (a): The gate level representation. (b): The quantum level representation. V and V+ represent the square root of NOT and its adjoint, respectively.

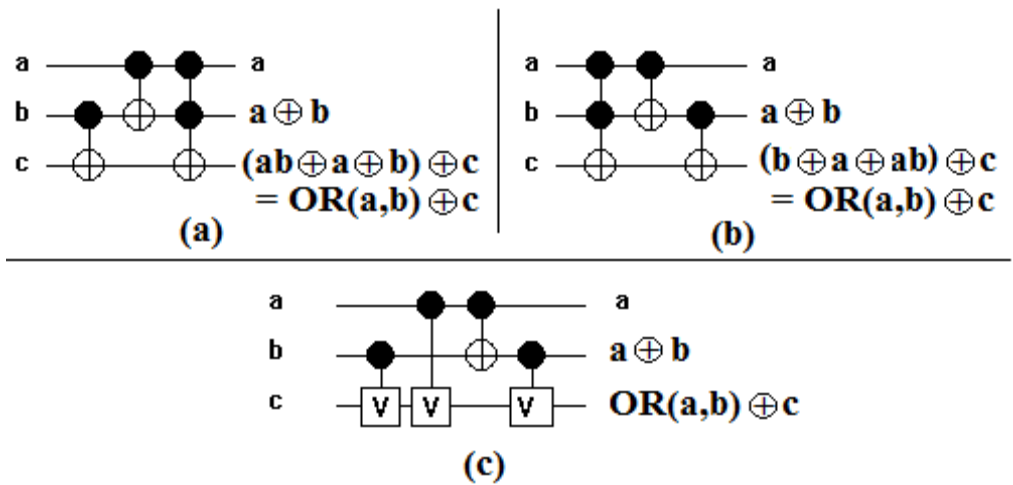

**Figure 4:** Or-Peres gate. (a) and (b): Gate level representations. (c): Quantum level representation.

### 2.3 RIMEP2 genetic operators

RIMEP2 is a Darwinian evolution driven system.

#### 1) Selection

Individuals for breeding are selected based on the "**tournament**" selection with a size of "**2**". The tournament is fitness-based. The degree of the competition (tournament size) is proportional to the population size.

#### 2) Crossover

RIMEP2 used the **uniform crossover** to evolve the whole set of the collected benchmarks. The number and the positions of the cuts are randomly chosen. No post-processing is needed to guaranty that functional and syntactically correct reversible circuits are evolved. This constitutes an important feature of RIMEP2. An example of the crossover operation is illustrated in Figure 5. The numbers appearing on the top the Figure 5 indicate the ordering of the sub-expressions in the chromosome. The selected cut points, in this example, are: 1, 3 and 6.

### 3) Mutation

Four types of mutation are considered, each randomly selected with respect to a threshold.

- Type 1: **mutate the operator** from the proposed library. If the new operator needs additional inputs then, the missing addresses (lines) should be filled. RIMEP2 assures that no line is used more than once for the same gate, thus, the fanout problem is avoided. See Figure 6.a.
- Type 2: **mutate an address (a gate input)**. The address should be replaced by another one which should be different and not already assigned to the same gate in order to maintain the circuit fanout free. See Figure 6.b.
- Type 3**: circular shifting of the addresses** (inputs) of a selected sub-expression (gate). See Figure 7.a.
- Type 4**: Exchanging the order of two randomly selected expressions (gates).** See Figure 7.b.

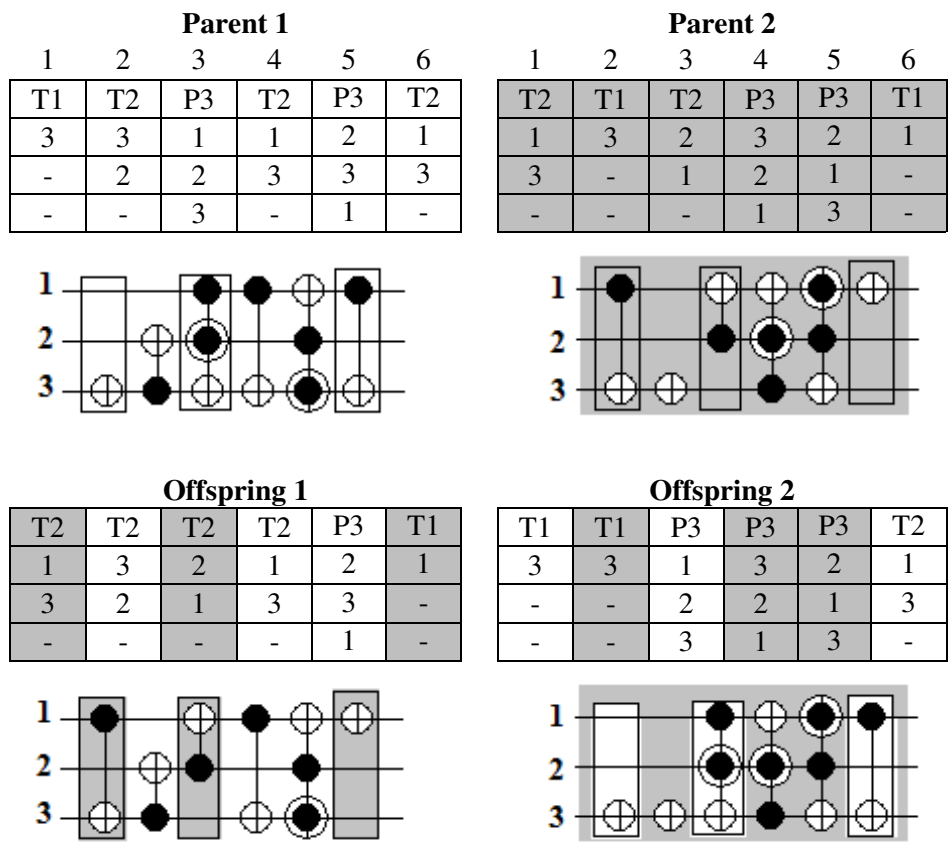

Figure 5: Crossover. The cut points are: 1, 3 and 6.

**Individual 1 before**                    **Individual 1 after**

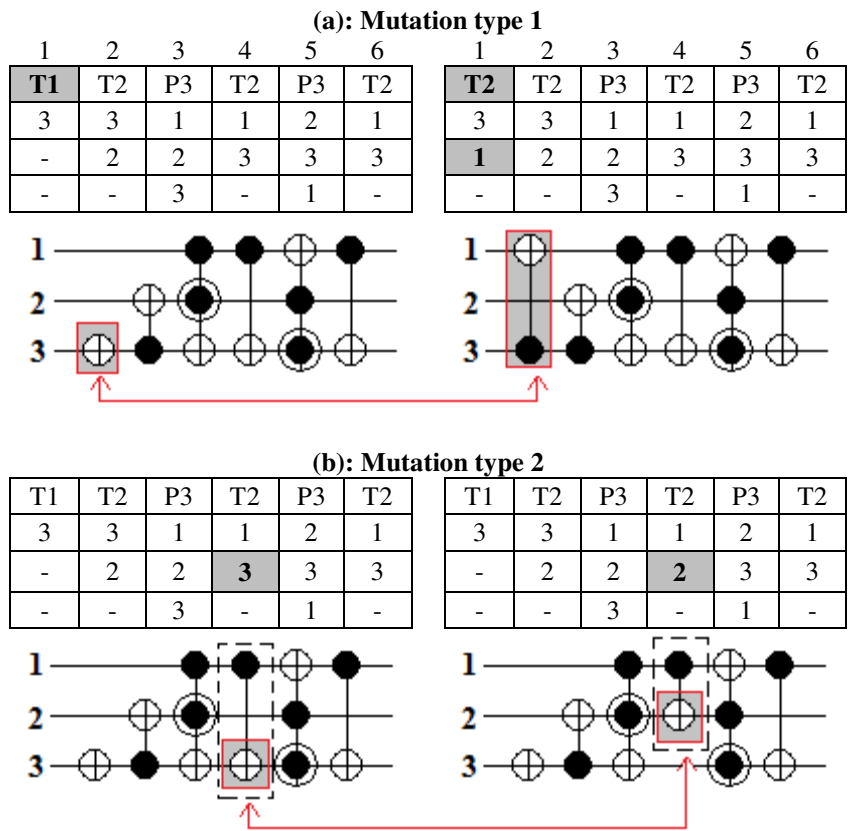

Figure 6 : The first and the second type of mutation.

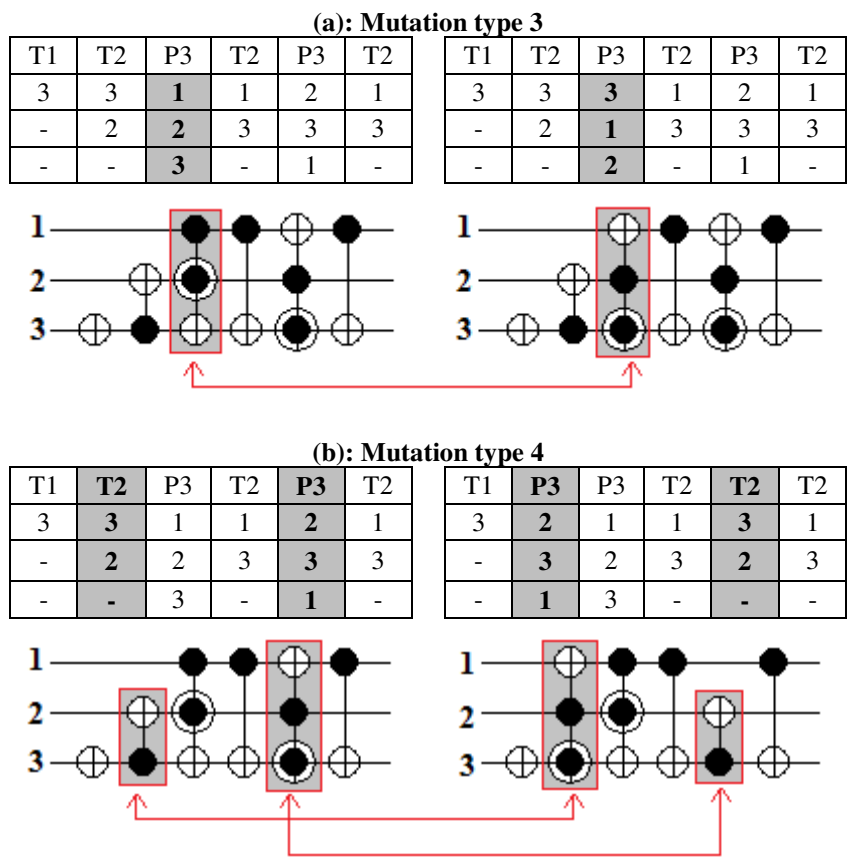

Figure 7: The third and the fourth type of mutation.

4) RIMEP2 flow chart

The flow chart of RIMEP2 is shown in Figure 8.

## 3. Some experiments

As we have mentioned above, we aimed to fully introduce the RIMEP2 system. In this section we show some of the evolved reversible circuits comparing them to their counterparts in the literature. In all the cases, the evolution was done from scratch without resorting to pre-existing libraries using a cluster of computers (LiDOng cluster of the TU Dortmund University, Germany, see [24] for more details about the total number of nodes, the CPU clock rate and other features. This support is here gladly acknowledged). This Report will serve as a reference work to the RIMEP2 system for some comparative studies to be investigated in the near future. Some of RIMEP2 parameters have been fixed based on a previous study [20]. Others have been determined according to some heuristics and variant experiments. The parameter setting is displayed in Table 4. Parameter fine-tuning may be done to improve the potential of RIMEP2 in discovering new optimal solutions for specified reversible function problems. Some of the values of Table 4 are given in ranges. Each benchmark has been evolved using an appropriate value from a range for each parameter. The results were quite encouraging and an interesting improvement in terms of quantum cost (QC) was noticed. A set of benchmarks was collected from different reference works. Their specifications are given in Table 5. We avoid listing the whole truth table for each; instead a decimal encoded specification will be shown. For example the decimal specification of the example shown in Table 2, is [7,1,4,3,0,2,6,5 ], where each decimal number corresponds to the value of $4x + 2y + z$ (in the notation of Table 2). Table 5 shows for each benchmark, its specification, its number of inputs and outputs (some additional inputs/outputs may be added for the purpose of reversibility), the best known QC in the literature and the RIMEP2 best found QC for this benchmark. The last column of the same table gives the percentage of improvement. We should emphasize that the comparison we are showing even presents a quite high improvement, is not the main aim of the current report, but then highlights RIMEP2. Additional RIMEP2 results (Table 6 and Table 7) are compared to the ones published in [22] and [23] where evolutionary algorithms have been used to evolve reversible circuits using the NTC library (T1, T2 and T3 corresponding to the NOT, C-NOT and Toffoli gates respectively).

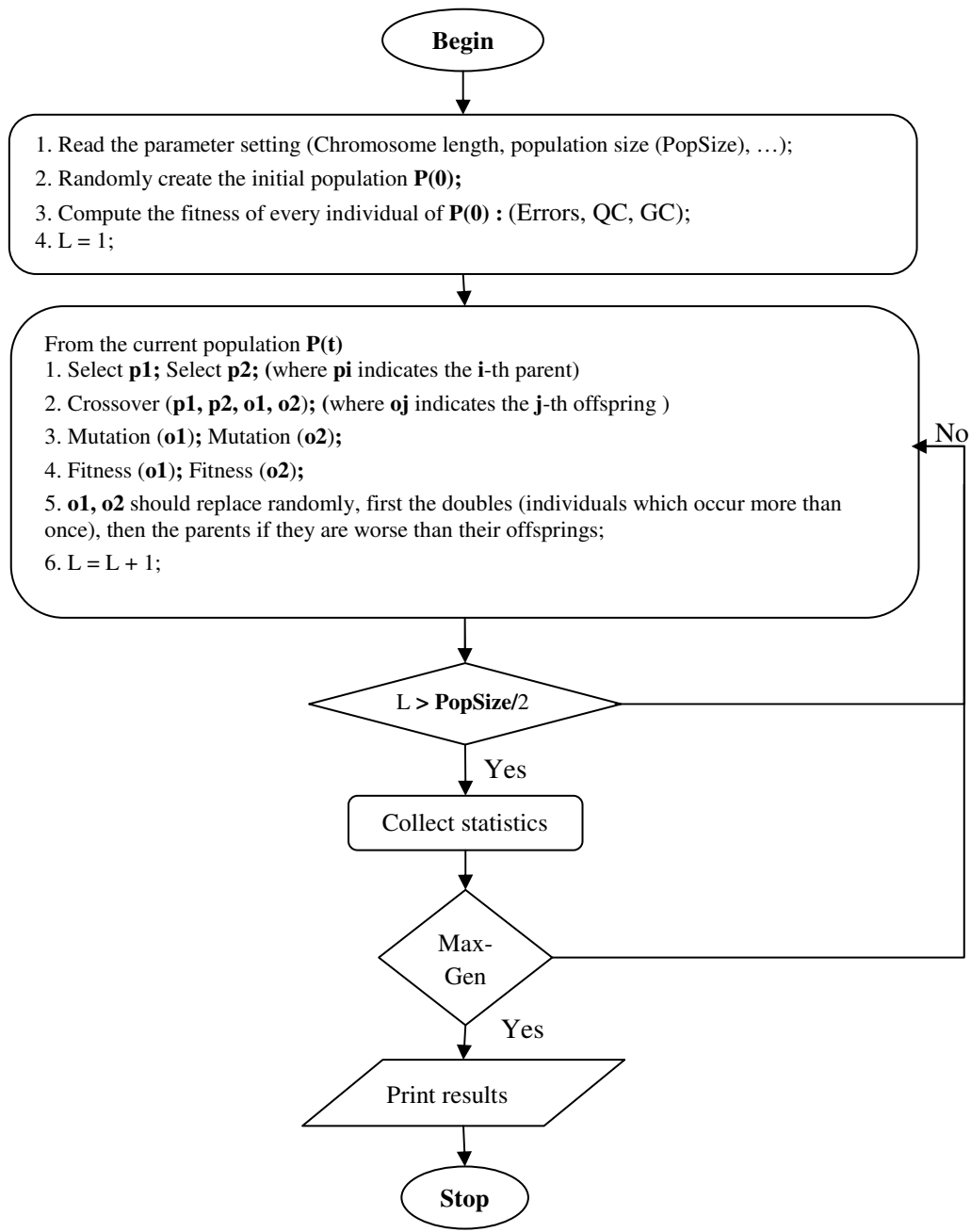

**Figure 8: RIMEP2 Flow chart.**

**Table 4: RIMEP2 parameter setting.**

| Parameter | Value |
|---|---|
| Crossover probability | 0.7 |
| Mutation probability | 0.01 |
| Chromosome length | 10-30 |
| Population size | 100-200 |
| Max-number of generations | 500-50,000 |
| Selection type | Tournament with size = 2 |
| Crossover | Uniform |
| Libraries (function sets) | {P3, T1, T2, T3}, {T1, T2, T3} |

**Table 5: Some RIMEP2 evolved benchmarks compared to their counterparts in the literature.**

| Benchmark name | Source | Decimal specification | Number of inputs and outputs (I/O) | RIMEP2 best QC | Best known QC | Improvement (%) | RIMEP2 CPU T in minutes |
|---|---|---|---|---|---|---|---|
| c17-204 | [21] | (0,1,0,1,0,1,0,0,3,3,3,3,3,0,0, 0,1,0,1,2,3,2,2,3,3,3,3,3,2,2) | 5/2 | 17 | 21 | 19.05 | 2.72 |
| hwb4 | [11] | (0,2,4,12,8,5,9,11,1,6,10,13,3, 14,7,15) | 4/4 | 18 | 19 | 5.26 | 0.33 |
| 4mod7 | [21] | (0,1,2,3,4,5,6,0,1,2,3,4,5,6,0,1) | 4/3 | 23 | 38 | 39.47 | 0.12 |
| 5mod5 | [9] | (1,0,0,0,0,1,0,0,0,0,1,0,0,0,0,1, 0,0,0,0,1,0,0,0,0,1,0,0,0,0,1,0) | 5/1 | 34 | 76 | 55.26 | 0.46 |
| majority | [21] | (0,0,0,0,0,0,0,1,0,0,0,1,0,1,1,1, 0,0,0,1,0,1,1,1,0,1,1,1,1,1,1,1) | 5/1 | 18 | 136 | 86.76 | 0.18 |
| mod5d1 | [21] | (0,1,3,2,5,4,7,6,9,8,10,11,13,1 2,15,14,17,16,19,18,20,21,23, 22,25,24,27,26,29,28,30,31) | 5/5 | 9 | 11 | 18.18 | 2.32 |
| mod5d2 | [21] | (21,6,3,0,9,26,15,12,13,14,27, 8,1,2,7,20,5,22,19,16,25,10, 31,28,29,30,11,24,17,18,23,4) | 5/5 | 14 | 16 | 12.50 | 0.85 |
| mod5mils | [21] | (3,2,0,1,6,7,4,5,8,9,11,10,12,1 3,14,15,18,19,16,17,23,22,20, 21,24,25,26,27,28,29,31,30) | 5/5 | 10 | 13 | 23.07 | 0.18 |
| cm42a | [21] | (511,1021,991,1023,895,1023, 1015,1023,767,1022,1007, 1023,959,1023,1019,1023) | 4/10 | 218 | 377 | 42.17 | 3.42 |
| cm82a | [21] | (0,2,2,1,4,6,6,5,4,6,6,5,2,1,1,3, 4,6,6,5,2,1,1,3,2,1,1,3,6,5,5,7) | 5/3 | 18 | 154 | 88.31 | 2.34 |
| dc1 | [21] | (119,3,62,31,75,93,109,19,127 ,91,0,0,0,0,0,0) | 4/7 | 86 | 416 | 79.32 | 6.34 |
| **Average** | | | | | | **42.66** | 1.75 |

Table 6: RIMEP2 Vs [22] and [23]. Benchmark specifications.

| # | Benchmark Decimal specification | Source | Number of inputs and outputs (I/O) |
|---|---|---|---|
| 1 | (1,0,3,2,5,7,4,6) | [22], [23] | 3/3 |
| 2 | (7,0,1,2,3,4,5,6) | [22], [23] | 3/3 |
| 3 | (0,1,2,3,4,6,5,7) | [22], [23] | 3/3 |
| 4 | (0,1,2,4,3,5,6,7) | [22], [23] | 3/3 |
| 5 | (1,2,3,4,5,6,7,0) | [22], [23] | 3/3 |
| 6 | (3,6,2,5,7,1,0,4) | [22], [23] | 3/3 |
| 7 | (1,2,7,5,6,3,0,4) | [22], [23] | 3/3 |
| 8 | (4,3,0,2,7,5,6,1) | [22], [23] | 3/3 |
| 9 | (7,5,2,4,6,1,0,3) | [22] | 3/3 |
| 10 | (0,7,4,3,2,5,1,6) | [22] | 3/3 |
| 11 | (7,1,4,3,0,2,6,5) | [22] | 3/3 |
| 12 | (1,2,3,4,5,6,7,8,9,10,11,12,13,14,15,0) | [22], [23] | 4/4 |
| 13 | (1,0,0,0,0,1,0,0,0,0,1,0,0,0,0,1) | [22] | 5/1 |
| 14 | (0,1,3,2,6,7,5,4,12,13,15,14,10,11,9,8,24,25,27,26,30,31,29,28, 20,21,23,22,18,19,17,16,48,49,51,50,54,55,53,52,60,61,63,62, 58,59,57,56,40,41,43,42,46,47,45,44,36,37,39,38,34,35,33,32) | [22] | 6/6 |
| 15 | (0,7,6,9,4,11,10,13,8,15,14,1,12,3,2,5) | [23] | 4/4 |

Table 7: Evolved benchmarks RIMEP2 Vs [22] and [23].

| Benchmark # | RIMEP2 | | Best known [22] | | Best known [23] | Improvement GC/QC (%) relative to the best between [22] and [23] |
|---|---|---|---|---|---|---|
| | best GC/QC | CPU T | best GC/QC | CPU T | best GC/QC | |
| 1 | 4 / 8 | 0.008 | 4 / 16 | <0.01 | 4 / 8 | 0 / 0 |
| 2 | 3 / 7 | 0.007 | 3 / 7 | <0.01 | 3 / 7 | 0 / 0 |
| 3 | **3 / 7** | 0.006 | 3 / 15 | <0.01 | 3 / 15 | **0 / 53** |
| 4 | 5 / 9 | 0.006 | 5 / 17 | 0.02 | 5 / 9 | 0 / 0 |
| 5 | **3 / 7** | 0.007 | 3 / 7 | <0.01 | 4 / 8 | 0 / 0 |
| 6 | **5 / 17** | 0.010 | 7 / 19 | 0.04 | 7 / 19 | **28 / 10** |
| 7 | **6 / 14** | 0.010 | 7 / 15 | 0.03 | 7 / 15 | **14 / 6** |
| 8 | **5 / 9** | 0.010 | 6 / 10 | 0.03 | 6 / 10 | **16 / 10** |
| 9 | **6 / 18** | 0.010 | 9 / 21 | 0.04 | Not reported | **33 / 14** |
| 10 | **3 / 7** | 0.010 | 5 / 9 | 0.02 | Not reported | 40 / 22 |
| 11 | **5 /13** | 0.010 | 6 / 14 | 0.03 | Not reported | **16 / 7** |
| 12 | 4 /20 | 0.030 | 6/26 | 0.02 | 4 / 20 | 0 / 0 |
| 13 | **5 / 9** | 0.090 | 5 / 55 | 0.05 | Not reported | **0 / 84** |
| 14 | 5 / 5 | 0.090 | 5 / 5 | 0.05 | Not reported | 0 / 0 |
| 15 | **5 / 9** | 0.090 | Not reported | | 6 / 10 | 16 / 10 |
| **Average Improvement GC/QC (%)** | | | | | | **9 / 13.73** |

## 4. Interpretation of the Results

Table 5 shows benchmarks selected, from two well known reversible circuits libraries [9] and [21], according to different input/output numbers where in some cases ancillary bits have to be added for reversibility purposes. The best results shown in column 6 of Table 5 were discovered using non-evolutionary techniques. Even the set of benchmarks was not large enough the results evolved by RIMEP2 (column 5 of the same table) display an improvement up to 88.31% with an average of 42.66 in the quantum cost. The CPU time is included shows a minimum of 0.12 min and a maximum of 6.34 min with an average of 1.75 min. One should point up that the CPU time is computed once the stop criterion of the program is met: the maximum number of generations. The solution may have been found before.

Additional experiments have been done to compare RIMEP2 to evolutionary methods for reversible design published in [22] and [23]. The benchmark specifications are shown in Table 6 and the results are given in Table 7 where quantum cost and gate count are compared. Even the shown reversible circuits were supposed to be already optimized by the method presented in [22] and [23], RIMEP2 was able evolve better solutions in most of the cases with improvements of up to 84% and 40% with averages of 9% and 13.73% for QC and GC respectively and matches the rest (we concluded that the optimum solutions in these cases were already encountered). Additionally, the CPU times for RIMEP2 and the method in [22] are shown. The comparison here is not emphasized because different hardware were used to design the presented benchmarks.

## 5. Conclusion and further work

An evolutionary system -RIMEP2- has been developed and tested for the design of a set of different reversible circuits. The evolution was done from scratch without resorting to a pre-existing library. The results show that among the 26 considered benchmarks, RIMEP2 outperformed the best published solutions for 20 of them and matched 6; reaching a quantum cost reduction up to 88.31% with an average of 42.66% in the quantum cost for non-evolutionary methods. Improvements up to 84% and 40% with averages of 9% and 13.73% for QC and GC respectively were achieved when RIMEP2 is compared to evolutionary methods described in [22] and [23].

RIMEP2 has been presented in this report as a promising method with a considerable potential for reversible circuit design. It will be considered as work reference for future studies based on this method.

Technical Reports ECSC of the European Centre for Soft Computing may be obtained upon email request (ecsc@claudio-moraga.eu).